\documentclass[12pt]{article}
\usepackage{amsfonts}

\usepackage{amsmath}
\usepackage{graphicx}

\input{tcilatex}

\begin{document}

\author{\ \  Diego Julio Cirilo Lombardo\thanks{%
diego77jcl@yahoo.com}\thanks{%
diego@thsun1.jinr.ru} \\
141980 Dubna, Russia\ \ \ \\
}
\title{The Newman-Janis Algorithm, Rotating Solutions and Einstein-Born-Infeld
Black Holes}
\maketitle

\begin{abstract}
A new metric is obtained by applying a complex coordinate transformation to
the static metric of the self-gravitating Born-Infeld monopole. The
behaviour of the new metric is typical of a rotating charged source, but
this source is not a spherically symmetric Born-Infeld monopole with
rotation. We show that the structure of the energy-momentum tensor obtained
with this new metric does not correspond to the typical structure of the
energy momentum tensor of Einstein-Born-Infeld theory induced by a rotating
spherically symmetric source. This also show, that the complex coordinate
transformations have the interpretation given by Newman and Janis only in
space-time solutions with linear sources.

PACS numbers:02.30.Dk, 04.20.Cv, 04.20.Jb
\end{abstract}

\section{Introduction}

\noindent \qquad There is a surprising connection between non-rotating and
rotating spacetime solutions of Einstein theory discovered firstly by Kerr
[1], and analyzed and interpreted by Newman and Janis [2] obtained by
applying a complex coordinate transformation. Using this method it was
possible to construct the Kerr solution from the Schwarzschild metric [1]
and also obtain from the Reissner-Nordstr\"{o}m metric its rotating
counterpart, the Kerr-Newman solution [3,4]. Quite recently, the rotating
Ba\~{n}ados-Teitelbaum-Zanelli (BTZ)\ black hole solution [5] and the
gravitational field of a rotating global monopole [6] were derived from
their non-rotating counterparts by using a complex transformation as was
pointed out by Janis and Newman.

In this paper, we perform a complex coordinate transformation into the
static spherically symmetric metric of a Born-Infeld monopole (in the form
discussed by Newman in [7]) to determine if the obtained new metric (that
has rotating and charged features) coincides with the metric obtained
directly of a Born-Infeld monopole with rotation. To this end, we compare
(in the same tetrad basis) the structure of the energy-momentum tensor from
the new metric (obtained with the Newman-Janis algorithm) with the typical
structure of the energy-momentum tensor of Born-Infeld from a rotating
charged source (obtained from the Einstein-Born-Infeld equations). One can
see that both the structures of the energy-momentum tensors are completely
\textit{different}. In this manner, we also show that the complex
transformation has the interpretation given by Newman et al.[2,3,4,7] only
in spacetime solutions with linear sources.

The plan of this paper is as follows: in Section 2 we give a short
introduction to the Born-Infeld theory: properties, principal features, and
static spherically symmetric solutions in this non-linear electrodynamic
theory. In Section 3 the Newman-Janis Algorithm (NJA), in the same manner
that was originally given by Newman and Janis [2,3,4], is applied to a
static spherically symmetric space-time of the Born-Infeld (electric)
monopole. Section 4 presents the analysis of the energy-momentum tensor and
the main result: the application of the NJA is correct \textit{only} when
the origin of the static ``seed'' metric is a ``linear'' source. Finally,
the conclusion and comments of the results are presented in Section 5.

\section{The Born-Infeld theory}

A non-linear theory of electrodynamics which keeps making appearances again
and again in many differents contexts within modern theoretical physics is
the Born-Infeld theory [11]. Among its many special properties is an exact
SO(2) electric-magnetic duality invariance. The Lagrangian density
describing Born-Infeld theory (in arbitrary space-time dimensions) is
\begin{equation}
\mathcal{L}_{BI}=\sqrt{g}L_{BI}=\frac{4\pi }{b^{2}}\left\{ \sqrt{g}-\sqrt{%
\left| \det (g_{\mu \nu }+bF_{\mu \nu })\right| }\right\}
\end{equation}
where $b$ is a fundamental parameter of the theory with field dimensions .
In open superstring theory [21], for example, loop calculations lead to this
Lagrangian with $b=2\pi \alpha
{\acute{}}%
$ $\left( \alpha
{\acute{}}%
\equiv \text{inverse of the string tension}\right) $\ . In four space-time
dimensions the determinant in (2) can be expanded out to give
\begin{equation*}
L_{BI}=\frac{4\pi }{b^{2}}\left\{ 1-\sqrt{1+\frac{1}{2}b^{2}F_{\mu \nu
}F^{\mu \nu }-\frac{1}{16}b^{4}\left( F_{\mu \nu }\widetilde{F}^{\mu \nu
}\right) ^{2}}\right\}
\end{equation*}
which coincides with the usual Maxwell Lagrangian in the weak field limit.

It is useful to define the second rank tensor $P^{\mu \nu }$ by
\begin{equation}
P^{\mu \nu }=-\frac{1}{2}\frac{\partial L_{BI}}{\partial F_{\mu \nu }}=\frac{%
F^{\mu \nu }-\frac{1}{4}b^{2}\left( F_{\mu \nu }\widetilde{F}^{\mu \nu
}\right) \,\widetilde{F}^{\mu \nu }}{\sqrt{1+\frac{1}{2}b^{2}F_{\mu \nu
}F^{\mu \nu }-\frac{1}{16}b^{4}\left( F_{\mu \nu }\widetilde{F}^{\mu \nu
}\right) ^{2}}}
\end{equation}
(so that $P^{\mu \nu }\approx F^{\mu \nu }$ for weak fields) satisfying the
electromagnetic equations of motion
\begin{equation}
\nabla _{\mu }P^{\mu \nu }=0
\end{equation}
which are highly non linear in $F_{\mu \nu }$. The energy-momentum tensor
can be written as
\begin{equation}
T_{\mu \nu }=\frac{1}{4\pi }\left\{ \frac{F_{\mu }^{\,\ \ \lambda }F_{\nu
\lambda }+\frac{1}{b^{2}}\left[ \sqrt{1+\frac{1}{2}b^{2}F_{\mu \nu }F^{\mu
\nu }-\frac{1}{16}b^{4}\left( F_{\mu \nu }\widetilde{F}^{\mu \nu }\right)
^{2}}-1-\frac{1}{2}b^{2}F_{\mu \nu }F^{\mu \nu }\right] g_{\mu \nu }}{\sqrt{%
1+\frac{1}{2}b^{2}F_{\mu \nu }F^{\mu \nu }-\frac{1}{16}b^{4}\left( F_{\mu
\nu }\widetilde{F}^{\mu \nu }\right) ^{2}}}\right\}
\end{equation}
Although it is by no means obvious, it can be verified that equations
(3)-(5) are invariant under electric-magnetic duality $F\longleftrightarrow
\ast G$. We can show that the SO$\left( 2\right) $ structure of the
Born-Infeld theory is more easily seen in quaternionic form [12,20]
\begin{equation*}
\frac{1}{R}\left( \sigma _{0}+i\sigma _{2}\overline{\mathbb{P}}\right) L=%
\mathbb{L}
\end{equation*}
\begin{equation*}
\frac{R}{\left( 1+\overline{\mathbb{P}}^{2}\right) }\left( \sigma
_{0}-i\sigma _{2}\overline{\mathbb{P}}\right) \mathbb{L}=L
\end{equation*}
\begin{equation*}
\overline{\mathbb{P}}\equiv \frac{\mathbb{P}}{b}
\end{equation*}
where we have been defined
\begin{equation*}
L=F-i\sigma _{2}\widetilde{F}
\end{equation*}
\begin{equation*}
\mathbb{L}=P-i\sigma _{2}\widetilde{P}
\end{equation*}
\begin{equation*}
R=\sqrt{1+\frac{1}{2}b^{2}F_{\mu \nu }F^{\mu \nu }-\frac{1}{16}b^{2}\left(
F_{\mu \nu }\widetilde{F}^{\mu \nu }\right) ^{2}}
\end{equation*}
the pseudoescalar of the electromagnetic tensor $F^{\mu \nu }$%
\begin{equation*}
\mathbb{P}=-\frac{1}{4}F_{\mu \nu }\widetilde{F}^{\mu \nu }
\end{equation*}
and $\sigma _{0}\,$, $\sigma _{2}$ the well know Pauli matrix.

In flat space, and for purely electric configurations, the Lagrangian (2)
reduces to
\begin{equation*}
L_{BI}=\frac{4\pi }{b^{2}}\left\{ 1-\sqrt{1+b^{2}\overrightarrow{E^{2}}}%
\right\}
\end{equation*}
so there is an upper bound on the electric field strength $\overrightarrow{E}
$
\begin{equation}
\left| \overrightarrow{E}\right| \leq \frac{1}{b}
\end{equation}
Due to a point charge the field is
\begin{equation*}
E_{r}=\frac{Q}{\sqrt{r^{4}+b^{2}Q^{2}}}
\end{equation*}
and so achieves the bound (6) at $r=0$. The total self-energy of the point
charge is thus
\begin{equation*}
\mathcal{E}=\frac{1}{4\pi }\int d^{3}x\,\ T_{00}=\frac{1}{4\pi }\int d^{3}x\,%
\frac{1}{b^{2}r^{2}}\left( \sqrt{r^{4}+b^{2}Q^{2}}-r^{2}\right)
\end{equation*}
Integrating by parts gives a standard elliptic integral
\begin{equation*}
\mathcal{E}=\frac{2Q}{3}\int_{0}^{\infty }\frac{dr}{\sqrt{r^{4}+b^{2}Q^{2}}}=%
\frac{\left( \pi Q\right) ^{3/2}}{3\sqrt{b}\left[ \Gamma \left( 3/4\right) %
\right] ^{2}}
\end{equation*}
which is finite (for simplicity, $Q$ and $b$ are taken to be positive here).
Thus, the Born-Infeld theory succeeded in its original goal of providing a
model for point charges with finite self-energy. Note that in the limit $%
b\rightarrow 0$, the Maxwell theory is reproduced and the self-energy
diverges.

Now consider static spherically symmetric black holes in this theory. Using
electric-magnetic duality, there is no loss of generality in considering
only electrically charged black holes. The solution is
\begin{equation*}
ds^{2}=-\left( 1-\frac{2GM\left( r\right) }{r}\right) dt^{2}+\left( 1-\frac{%
2GM\left( r\right) }{r}\right) ^{-1}dr^{2}+r^{2}\left( d\theta ^{2}+\sin
^{2}\theta \ d\varphi ^{2}\right)
\end{equation*}
\begin{equation*}
\begin{array}{ccc}
P_{tr}=\frac{Q}{r^{2}} &  & F_{tr}=\frac{Q}{\sqrt{r^{4}+b^{2}Q^{2}}}
\end{array}
\end{equation*}
where the function $M(r)$ satisfies
\begin{equation}
M%
{\acute{}}%
\left( r\right) =\frac{1}{b^{2}}\left( \sqrt{r^{4}+b^{2}Q^{2}}-r^{2}\right)
\end{equation}
and
\'{}%
denotes differentiation with respect to $r$. The mass M is given by
\begin{equation*}
\text{M=}\lim_{r\rightarrow \infty }M\left( r\right)
\end{equation*}
and hence
\begin{equation*}
M\left( r\right) =\text{M}-\frac{1}{b^{2}}\int_{r}^{\infty }dx\,\left( \sqrt{%
x^{4}+b^{2}Q^{2}}-x^{2}\right)
\end{equation*}
which is a monotonically increasing function of $r$. The horizons are given
by the roots of the equation $r=2M(r)$ and so the number of horizons will be
determined by $M(0)$ and $M%
{\acute{}}%
(0)$; $M(0)$ depends on the self-energy of the electromagnetic field, the
integral being the same as for the point charge in flat space
\begin{equation*}
M\left( 0\right) =\text{M}-\frac{\left( \pi Q\right) ^{3/2}}{3\sqrt{b}\left[
\Gamma \left( 3/4\right) \right] ^{2}}=\text{M}-\mathcal{E}
\end{equation*}
and so $M(0)$ may be interpreted as the binding energy. From (7) one has
\begin{equation*}
M%
{\acute{}}%
\left( 0\right) =\frac{Q}{b}
\end{equation*}
For $M(0)>0$ there is precisely one non-degenerate horizon. If $M(0)=0,$
then there is one non-degenerate horizon for $Q>b/2$ and none otherwise. The
case $M(0)<0$ is similar to Reissner-Nordstr\"{o}m, with either no horizons,
one degenerate horizon or two non-degenerate horizons, depending on the
relative magnitudes of $M$, $Q$, and $b$. Note that the
Reissner-Nordstr\"{o}m solution is recovered in the limit $b\rightarrow 0$
in which $M(0)\rightarrow -\infty $.

\section{The NJA\ and the rotating charged non linear solution}

The static solution of self-gravitating monopole in the electromagnetic
Born-Infeld theory was investigated by B. Hoffmann [8]. He showed that the
gravitational field is described by a static and spherically symmetric
metric with a non-linear electromagnetic source. The space-time metric
produced by the static charged nonlinear source is given by [8,9]
\begin{equation}
ds^{2}=-\left( 1-\frac{2GM}{r}+\frac{Q^{2}\left( r\right) }{r^{2}}\right)
dt^{2}+\left( 1-\frac{2GM}{r}+\frac{Q^{2}\left( r\right) }{r^{2}}\right)
^{-1}dr^{2}+r^{2}\left( d\theta ^{2}+\sin ^{2}\theta \ d\varphi ^{2}\right)
\end{equation}
where
\begin{equation*}
c=1
\end{equation*}
\begin{equation*}
M\equiv M_{Schwarzschild}
\end{equation*}
\begin{align}
Q^{2}\left( r\right) & \equiv Q^{2}\left\{ 2\left[ \frac{\overline{r}^{4}}{3}%
-\frac{\sqrt{1+\overline{r}^{4}}}{3}\overline{r}^{2}+\frac{2}{3}\overline{r}%
\left( -1\right) ^{\frac{1}{4}}F\left[ Arc\sin \left[ \left( -1\right) ^{%
\frac{3}{4}}\overline{r}\right] ,-1\right] \right] \right\} \\
& =Q^{2}\left\{ 2\left[ \frac{\overline{r}^{4}}{3}-\frac{\sqrt{1+\overline{r}%
^{4}}}{3}\overline{r}^{2}-\frac{2}{3}\overline{r}^{2}\ \ _{2}F_{1}\left[
1/4,1/2,5/4,-\overline{r}^{4}\right] \right] \right\}  \notag
\end{align}
Where we have been utilized the well-know relation between the incomplete
elliptic function of first class $F$ and the Gauss hypergeometric[10]
function $_{2}F_{1}$%
\begin{equation}
-\left( -1\right) ^{\frac{1}{4}}F\left[ Arc\sin \left[ \left( -1\right) ^{%
\frac{3}{4}}\overline{r}\right] ,-1\right] =\overline{r}\ \ _{2}F_{1}\left[
1/4,1/2,5/4,-\overline{r}^{4}\right]
\end{equation}
and
\begin{equation*}
\overline{r}\equiv \frac{r}{r_{0}}
\end{equation*}
where $r_{0}$ is the Born-Infeld radius related to the non-linear parameter $%
b$ in the following form [11]
\begin{equation*}
b=\frac{Q}{r_{0}^{2}}.
\end{equation*}
To apply the NJA\ to the static Born-Infeld monopole, we follow the steps
that were originally given by Newman and Janis. In the Eddington-Finkelstein
type coordinates this metric can be rewritten as
\begin{equation}
ds^{2}=-\left( \frac{\Delta }{r^{2}}\right) du^{2}-2du\ dr+r^{2}\left(
d\theta ^{2}+\sin ^{2}\theta \ d\varphi ^{2}\right)
\end{equation}
where the new variable $u$ is defined by
\begin{equation}
u=t-r-f\left( r\right)
\end{equation}
and
\begin{equation}
\Delta =r^{2}-2GMr+Q^{2}\left( r\right)
\end{equation}
From (11) we can read the contravariant component of the metric, namely,
\begin{equation}
\begin{array}{ccccc}
g^{00}=0; & g^{01}=-1; & g^{11}=\left( 1-\frac{2GM}{r}+\frac{Q^{2}\left(
r\right) }{r^{2}}\right) ; & g^{22}=\frac{1}{r^{2}}; & g^{33}=\frac{1}{%
r^{2}\sin ^{2}\theta }
\end{array}
\end{equation}
which can be written in a different form
\begin{equation}
g^{\mu \nu }=-l^{\mu }n^{v}-l^{\nu }n^{\mu }+m^{\mu }\overline{m}^{\nu
}+m^{\nu }\overline{m}^{\mu }
\end{equation}
where the null tetrad vectors are
\begin{equation}
\begin{array}{ccc}
l^{\mu }=\delta _{1}^{\mu } &  & n^{\mu }=\delta _{0}^{\mu }-\frac{1}{2}%
\left( 1-\frac{2GM}{r}+\frac{Q^{2}\left( r\right) }{r^{2}}\right) \delta
_{1}^{\mu } \\
&  &  \\
m^{\mu }=\frac{1}{\sqrt{2}r}\left( \delta _{2}^{\mu }+\frac{i}{\sin
^{{}}\theta }\delta _{3}^{\mu }\right) &  & \overline{m}^{\mu }=\frac{1}{%
\sqrt{2}r}\left( \delta _{2}^{\mu }-\frac{i}{\sin ^{{}}\theta }\delta
_{3}^{\mu }\right)
\end{array}
\end{equation}
with $\overline{m}^{\mu }$ being the complex conjugate of $m^{\mu }$.

Now, following Newman et al. [2,3,4,7], the radial coordinate is allowed to
be complex and the tetrad null vectors can be rewritten as
\begin{equation}
\begin{array}{ccc}
l^{\mu }=\delta _{1}^{\mu } &  & n^{\mu }=\delta _{0}^{\mu }-\frac{1}{2}%
\left[ 1-GM\left( \frac{1}{r}+\frac{1}{r^{\ast }}\right) +\frac{Q^{2}\left(
r,r^{\ast }\right) }{rr^{\ast }}\right] \delta _{1}^{\mu } \\
m^{\mu }=\frac{1}{\sqrt{2}r^{\ast }}\left( \delta _{2}^{\mu }+\frac{i}{\sin
^{{}}\theta }\delta _{3}^{\mu }\right) &  & \overline{m}^{\mu }=\frac{1}{%
\sqrt{2}r}\left( \delta _{2}^{\mu }-\frac{i}{\sin ^{{}}\theta }\delta
_{3}^{\mu }\right)
\end{array}
\end{equation}
where $r^{\ast }$ is the complex conjugate of $r$ and
\begin{equation}
\frac{Q^{2}\left( r,r^{\ast }\right) }{rr^{\ast }}=\frac{Q^{2}}{rr^{\ast }}%
\left\{ 2\left[ \frac{\overline{\rho }^{4}}{3}-\frac{\sqrt{1+\overline{\rho }%
^{4}}}{3}\overline{\rho }^{2}-\frac{2}{3}\overline{\rho }^{2}\ \ _{2}F_{1}%
\left[ 1/4,1/2,5/4,-\overline{\rho }^{4}\right] \right] \right\}
\end{equation}
with
\begin{equation*}
\overline{\rho }^{2}\equiv \frac{rr^{\ast }}{r_{0}^{2}}
\end{equation*}
The next step is to perform the complex coordinate transformation
\begin{equation}
\begin{array}{ccc}
\widetilde{u}=u-ia\cos \theta , &  & \widetilde{r}=r+ia\cos \theta \\
\widetilde{\theta }=\theta , & and & \widetilde{\varphi }=\varphi
\end{array}
\end{equation}
on $l^{\mu }$ $n^{\mu }$ and $m^{\mu }$. Replacing (19) in (17) we have
\begin{equation*}
\begin{array}{ccc}
\widetilde{l}^{\mu }=\delta _{1}^{\mu }\ \ \ \ \ \ \ \ \ \ \  &  &
\widetilde{n}^{\mu }=\delta _{0}^{\mu }-\frac{1}{2}\left[ 1-GM\left( \frac{%
\widetilde{r}}{\widetilde{r}^{2}+a^{2}\cos ^{2}\widetilde{\theta }}\right) +%
\frac{Q^{2}\left( r,r^{\ast }\right) }{\widetilde{r}^{2}+a^{2}\cos ^{2}%
\widetilde{\theta }}\right] \delta _{1}^{\mu }
\end{array}
\end{equation*}
\begin{equation*}
\widetilde{m}^{\mu }=\frac{1}{\sqrt{2}(\widetilde{r}+ia\cos \widetilde{%
\theta })}\left( ia\sin \widetilde{\theta }\left( \delta _{0}^{\mu }-\delta
_{1}^{\mu }\right) +\delta _{2}^{\mu }+\frac{i}{\sin ^{{}}\widetilde{\theta }%
}\delta _{3}^{\mu }\right)
\end{equation*}
\begin{equation}
\widetilde{\overline{m}}^{\mu }=\frac{1}{\sqrt{2}(\widetilde{r}-ia\cos
\widetilde{\theta })}\left( -ia\sin \widetilde{\theta }\left( \delta
_{0}^{\mu }-\delta _{1}^{\mu }\right) +\delta _{2}^{\mu }-\frac{i}{\sin ^{{}}%
\widetilde{\theta }}\delta _{3}^{\mu }\right)
\end{equation}
where $\widetilde{\overline{m}}^{\mu }$ is defined as the complex conjugate
of $\widetilde{m}^{\mu }.$

\bigskip The complex transformation which has been made to the vectors of
the tetrad null, finally gives
\begin{equation}
\widetilde{g}^{\mu \nu }=-\widetilde{l}^{\mu }\widetilde{n}^{v}-\widetilde{l}%
^{\nu }\widetilde{n}^{\mu }+\widetilde{m}^{\mu }\ \widetilde{\overline{m}}%
^{\nu }+\widetilde{m}^{\nu }\widetilde{\overline{m}}^{\mu }
\end{equation}
Now, looking at the metric (21), the question is the following: `` is
expression (21) the metric of a rotating spherically simmetric Born-Infeld
monopole with angular momentum per unit of mass $a$?''. From (21) we can
read off the contravariant and covariant components of the metric. The
covariant components of the metric (21) in terms of the coordinates $\left(
\widetilde{u},\widetilde{r},\widetilde{\theta },\widetilde{\varphi }\right) $
read
\begin{equation}
\widetilde{g}_{\mu \nu }=\left[
\begin{array}{cccc}
\frac{a^{2}\sin ^{2}\widetilde{\theta }-\Delta _{new}}{\widetilde{r}%
^{2}+a^{2}\cos ^{2}\widetilde{\theta }} & -1 & 0 & \frac{a\sin ^{2}%
\widetilde{\theta }\left[ \Delta _{new}-\left( \widetilde{r}%
^{2}+a^{2}\right) \right] }{\widetilde{r}^{2}+a^{2}\cos ^{2}\widetilde{%
\theta }} \\
-1 & 0 & 0 & a\sin ^{2}\widetilde{\theta } \\
0 & 0 & \widetilde{r}^{2}+a^{2}\cos ^{2}\widetilde{\theta } & 0 \\
\frac{a\sin ^{2}\widetilde{\theta }\left[ \Delta _{new}-\left( \widetilde{r}%
^{2}+a^{2}\right) \right] }{\widetilde{r}^{2}+a^{2}\cos ^{2}\widetilde{%
\theta }} & a\sin ^{2}\widetilde{\theta } & 0 & \frac{\sin ^{2}\widetilde{%
\theta }\left[ \left( \widetilde{r}^{2}+a^{2}\right) ^{2}-\Delta
_{new}a^{2}\sin ^{2}\widetilde{\theta }\right] }{\widetilde{r}^{2}+a^{2}\cos
^{2}\widetilde{\theta }}
\end{array}
\right]
\end{equation}
\begin{equation}
\Delta _{new}=r^{2}-2GMr+Q^{2}\left( r,r^{\ast }\right) +a^{2}
\end{equation}
The metric given in (22) is not in the appropiate form to investigate if it
corresponds to a rotating charged source. One knows that the expansion in
power series up to the first order in $r/a$ (slowly rotating: $r/a<<1$) of
the expression within the brackets of the term of charge into $\Delta _{new}$
is [10,11,12]:
\begin{align}
\frac{Q^{2}\left( r,r^{\ast }\right) }{rr^{\ast }}& =\frac{Q^{2}}{rr^{\ast }}%
\left\{ 2\left[ \frac{\overline{\rho }^{4}}{3}-\frac{\sqrt{1+\overline{\rho }%
^{4}}}{3}\overline{\rho }^{2}-\frac{2}{3}\overline{\rho }^{2}\ \ _{2}F_{1}%
\left[ 1/4,1/2,5/4,-\overline{\rho }^{4}\right] \right] \right\} \simeq \\
& \simeq \frac{Q^{2}}{rr^{\ast }}\left\{ 2\left[ \frac{\overline{r}^{4}}{3}-%
\frac{\sqrt{1+\overline{r}^{4}}}{3}\overline{r}^{2}-\frac{2}{3}\overline{r}%
^{2}\ \ _{2}F_{1}\left[ 1/4,1/2,5/4,-\overline{r}^{4}\right] \right] \right\}
\notag
\end{align}
one can see from the last expansion that $\Delta _{new}$ is approximately
\begin{equation}
\Delta _{new}\simeq r^{2}-2GMr+Q^{2}\left( r\right) +a^{2}
\end{equation}
The condition that the function will depend only on the radius $r$ , in
addition to simplify the study of this type of complex transformations, is
need for the next step: to pass from the Kerr to the Boyer and Lindquist
coordinates [13] in which the rotating feature of the geometry is easy to
see (preferable frame). We might want to further transform it into one
written in the Boyer and Lindquist coordinates, with the following
transformations[13,14], which leaves invariant the block $u,\varphi $ of \ $%
\widetilde{g}_{\mu \nu }$(dropped subname to $\Delta _{new}$)$:$%
\begin{align}
d\widetilde{u}& =dt-\left( \frac{r^{2}+a^{2}}{\Delta }\right) dr \\
d\widetilde{\varphi }& =d\varphi -\frac{a}{\Delta }\ dr  \notag
\end{align}
Using these transformations, the metric given in (22) turns into
\begin{equation}
g_{\mu \nu }=\left[
\begin{array}{cccc}
\frac{a^{2}\sin ^{2}\theta -\Delta }{\rho ^{2}} & 0 & 0 & \frac{a\sin
^{2}\theta \left[ \Delta -\left( r^{2}+a^{2}\right) \right] }{\rho ^{2}} \\
0 & \frac{\rho ^{2}}{\Delta } & 0 & 0 \\
0 & 0 & \rho ^{2} & 0 \\
\frac{a\sin ^{2}\theta \left[ \Delta -\left( r^{2}+a^{2}\right) \right] }{%
\rho ^{2}} & 0 & 0 & \frac{\sin ^{2}\theta \left[ \left( r^{2}+a^{2}\right)
^{2}-\Delta a^{2}\sin ^{2}\theta \right] }{\rho ^{2}}
\end{array}
\right]
\end{equation}
Restoring explicitly the value of the $\Delta $ function (in the limit
indicated previously)
\begin{equation}
\Delta \simeq r^{2}-2GMr+Q^{2}\left( r\right) +a^{2}
\end{equation}
We get the metric of static Born-Infeld monopole obtained by Hoffmann in [8]
if we set $a=0$ in the metric (27). In the same way, if we set $Q=0$ (with $%
a\neq 0$),we get the Kerr solution in the Boyer-Lindquist coordinates
[13,14]. One can see that the metric (27) has the behaviour of a metric
produced from a rotating charged source, but we show in the next section
that the source of (27) is not a Born-Infeld monopole with rotation.

\section{Analysis of the energy-momentum tensor:}

One can see that applying complex transformations to the tetrad $l,n,m$
(complex), we can pass from the static system to a stationary solution of
Einstein equations. Will this new stationary solution be able to represent
the metric of a rotating spherical Born-Infeld monopole?. To do this, we
have to compare the structure of energy-momentum tensor from the new metric
with the structure of the energy-momentum tensor of Born-Infeld fields
obtained directly from a rotating charged source. Both the energy-momentum
tensors will be in the same basis. The new metric obtained by applying the
complex algorithm to the static problem analyzed by Hoffmann corresponds, in
the above aproximation, to:
\begin{equation}
\Delta _{rot}\simeq r^{2}+a^{2}-2\overline{M}\left( r\right) \ r\,\ \ \ \ \
\ \ \ \ \ \ \ \ \ \ (Geometrized\,Units,\,\text{[14]})
\end{equation}
where the function $\overline{M}(r)$ is:
\begin{equation}
\overline{M}\left( r\right) \equiv M_{s}-\frac{Q^{2}}{r}\left\{ \frac{%
\overline{r}^{4}}{3}-\frac{\sqrt{1+\overline{r}^{4}}}{3}\overline{r}^{2}+%
\frac{2}{3}\overline{r}\left( -1\right) ^{\frac{1}{4}}F\left[ Arc\sin \left[
\left( -1\right) ^{\frac{3}{4}}\overline{r}\right] ,-1\right] \right\}
\end{equation}
and
\begin{equation*}
M_{s}=Schwarzschild\ mass\left( ADM\right)
\end{equation*}
\begin{equation*}
\overline{r}\equiv \frac{r}{r_{0}}
\end{equation*}
where we put by convenience
\begin{equation}
\Delta _{rot}=r^{2}+a^{2}-2\ f\left( r\right)
\end{equation}
where\footnote{%
Note that for the Kerr-Newman black hole: \ $f\left( r\right) =mr-\frac{Q^{2}%
}{2}$}
\begin{equation}
f\left( r\right) =\overline{M}\left( r\right) \ r
\end{equation}
Looking at the metric
\begin{equation}
g_{\mu \nu }=n_{\mu }n_{\nu }+m_{\mu }m_{\nu }+l_{\mu }l_{\nu }-u_{\mu
}u_{\nu }
\end{equation}
where the vectors of the tetrad are
\begin{equation}
\begin{array}{l}
n_{\mu }=\sqrt{\Sigma }\left( 0,0,1,0\right) \\
l_{\mu }=\sqrt{\frac{\Sigma }{\Delta }}\left( 0,1,0,0\right) \\
u_{\mu }=-\sqrt{\frac{\Delta }{\Sigma }}\left( 1,0,0,-a\sin ^{2}\theta
\right) \\
m_{\mu }=\frac{\sin \theta }{\sqrt{\Sigma }}\left( a,0,0,-\left(
r^{2}+a^{2}\right) \right) \ \ ,
\end{array}
\end{equation}
$t,r,\theta ,\varphi $ being the Boyer and Lindquist coordinates. If we now
put in the Einstein equations [18] everything we have analyzed, in function
of $f\left( r\right) $ we can define [15,16,17]
\begin{equation*}
\rho ^{2}\equiv \Sigma
\end{equation*}
\begin{equation*}
D\equiv -\frac{f\ "\left( r\right) ^{2}}{\rho ^{2}}
\end{equation*}
\begin{equation}
G\equiv \frac{f\
{\acute{}}%
\left( r\right) r-f\left( r\right) }{\rho ^{4}}=\frac{r^{2}}{\rho ^{4}}%
\partial _{r}\left( \frac{f\left( r\right) }{r}\right)
\end{equation}
then the energy-momentum tensor takes the form [15,16,17]
\begin{equation}
T_{\mu \nu }=\frac{1}{8\pi }\left[ \left( D+2G\right) g_{\mu \nu }-\left(
D+4G\right) \left( l_{\mu }l_{\nu }-u_{\mu }u_{\nu }\right) \right]
\end{equation}
\begin{equation}
\Longrightarrow T_{\mu \nu }=\frac{1}{8\pi }\left[
\begin{array}{cccc}
2G & 0 & 0 & 0 \\
0 & -2G & 0 & 0 \\
0 & 0 & 2G+D & 0 \\
0 & 0 & 0 & 2G+D
\end{array}
\right]
\end{equation}
One can see that the $T_{\mu \nu }$ takes the same form as $T_{\mu \nu }$ of
an anisotropic fluid, being
\begin{equation}
\begin{array}{ccc}
\rho =\frac{1}{8\pi }2G, & p_{rad}=-\frac{1}{8\pi }2G, & p_{tg}=\frac{1}{%
8\pi }\left( 2G+D\right)
\end{array}
\end{equation}
where
\begin{equation}
D=-f\ "(r)\frac{1}{\rho ^{2}}=\frac{2Q^{2}}{\rho ^{2}r_{o}^{2}}\left[ 2%
\overline{r}^{2}-\frac{1+2\overline{r}^{4}}{\sqrt{1+\overline{r}^{4}}}\right]
\end{equation}
\begin{equation*}
G=\left( f\,%
{\acute{}}%
(r)r-f\right) \frac{1}{\rho ^{4}}=\frac{\overline{r}^{2}Q^{2}}{\rho ^{4}}%
\left[ \sqrt{1+\overline{r}^{4}}-\overline{r}^{2}\right]
\end{equation*}
\begin{equation*}
\text{\ (being }Q^{2}=b^{2}r_{0}^{4}\text{)}
\end{equation*}
From (35)-(39) we finally obtain
\begin{equation}
T_{00}=\frac{b^{2}}{4\pi }\left[ \frac{r^{2}}{\rho ^{4}}\left( \sqrt{%
r_{0}^{4}+r^{4}}-r^{2}\right) \right] =-T_{22}
\end{equation}
\begin{equation*}
T_{33}=\frac{b^{2}}{4\pi }\left\{ \frac{r^{2}}{\rho ^{4}}\left( \sqrt{%
r_{0}^{4}+r^{4}}-r^{2}\right) +\frac{1}{\rho ^{2}}\left[ 2r^{2}-\frac{%
r_{0}^{4}+2r^{4}}{\sqrt{r_{0}^{4}+r^{4}}}\right] \right\} =T_{11}
\end{equation*}
On the other hand, the structure of the metric energy-momentum tensor of
Born-Infeld, which was constructed from the electromagnetic fields of a
spherically \ symmetric source with rotation, in the same tetrad (34) is
[12, Appendix]:
\begin{equation*}
-T_{00}=T_{22}=\frac{b^{2}}{4\pi }\left( 1-\widetilde{u}\right)
\end{equation*}
\begin{equation}
T_{11}=T_{33}=\frac{b^{2}}{4\pi }\left( 1-\widetilde{u}^{-1}\right)
\end{equation}
with :
\begin{equation*}
\ \widetilde{u}\equiv \sqrt{\frac{b^{2}+F_{31}^{2}}{b^{2}-F_{20}^{2}}}
\end{equation*}
Note that $\widetilde{u}$ depends on the invariants of the electromagnetic
tensor $F_{ab}$ , and in the tetrad (34)(orthonormal frame) the relations
between the components of the energy-momentum tensor of Born-Infeld are
always (41). From here we see that
\begin{equation}
\widetilde{u}=\frac{T_{00}}{T_{33}}=\frac{r^{4}}{\rho ^{4}}\left( \sqrt{%
\frac{r_{0}^{4}}{r^{4}}+1}-1\right) \frac{1}{\left[ \frac{r^{4}}{\rho ^{4}}%
\left( \sqrt{\frac{r_{0}^{4}}{r^{4}}+1}-1\right) +\frac{r^{2}}{\rho ^{2}}%
\left( 2-\frac{\frac{r_{0}^{4}}{r^{4}}+2}{\sqrt{\frac{r_{0}^{4}}{r^{4}}+1}}%
\right) \right] }
\end{equation}
Let us note that if $\rho =r\Rightarrow \widetilde{u}_{\left( r=\rho \right)
}=\sqrt{\frac{r_{0}^{4}}{r^{4}}+1}$ $\equiv \widetilde{u}_{est}$ coincides
with $\widetilde{u}$ of the static case analyzed by Hoffmann [8,9] and the
typical structure of the Born-Infeld energy-momentum tensor (41) is
automatically satisfied with $\widetilde{u}_{est}$ given above, but for the
new case (obtained by means of the Newman-Janis algorithm), it is impossible
to reconstruct the energy-momentum tensor keeping the structure (41). For
example,
\begin{equation}
\Rightarrow T_{00}\neq \frac{b^{2}}{4\pi }\left( \widetilde{u}-1\right)
\end{equation}
with the function $\widetilde{u}$ given by expression (42). In this manner
we have demonstrated that the new metric (27) originating from the complex
transformation, does not correspond to the metric of a rotating spherically
symmetric Born-Infeld monopole.

\section{Conclusions:}

In this work, we have show that complex transformations, in the form pointed
out by Newman and Janis, for to obtain rotating solutions from the static
counterparts are only possible if the theory (source of the curvature) is
linear.

The limit as $a\rightarrow 0$ is still correct from the point of view of the
obtained solution, but the structure of the energy-momentum tensor for the
metric obtained by the complex Newman's transformation is completely
different to the structure of the energy-momentum tensor obtained directly
from the Einstein-Born-Infeld equations for the space-time of a spherical
monopole with rotation, both structures are in the same rotating frame
(basis tetrad) .

We have to analyze in more detail this type of complex transformations for
to see if it is possible to modify them for to include non-linear
electromagnetic sources in a next paper. The analysis of the Hamilton-Jacobi
equations[18] shows to us that if $\Delta =\Delta \lbrack f\left( r,\theta
\right) ]$ with $f\left( r,\theta \right) $ an trascendental function of $r$
and $\theta ,$ the dynamic problem is non-integrable (separability condition
for a Kerr type problem).

It is not difficult to see that the absence of null congruences in the
geometry, thus indicating that the metric is no longer\ Type D of the Petrov
classification [3,4,19].\bigskip

\section{Appendix:}

We will find the components of the energy-momentum tensor of Born-Infeld in
the rotating system. For this purpose, we have to take the metric
energy-momentum tensor

\begin{equation}
\frac{1}{\sqrt{\left| g\right| }}\frac{\delta \left( \sqrt{\left| g\right| }%
\mathcal{L}_{BI}\right) }{\delta g^{\mu \nu }}=-\ \frac{1}{2}\ T_{\mu \nu }\
\end{equation}
which is obtained by means of the standard variational procedure
(Einstein-Born-Infeld equations)[8,11,12]. In the symmetrized form[9] $%
T_{b}^{a}$ for a tetrad is
\begin{equation}
T_{\ \ b}^{a}=\delta _{\ \ b}^{a}\mathcal{L}_{BI}\mathcal{-}\frac{\partial
\mathcal{L}_{BI}}{\partial S}F_{\ \ \ l}^{a}F_{\ \ b}^{l}-\frac{\partial
\mathcal{L}_{BI}}{\partial P}F_{\ l}^{a}\widetilde{F}_{\ b}^{l}
\end{equation}
where the invariants (scalar and pseudoscalar) of the electromagnetic fields
are
\begin{equation}
S\equiv -\frac{1}{4}F_{ab}F^{ab}=\mathcal{L}_{M}
\end{equation}
\begin{equation}
P\equiv -\frac{1}{4}F_{ab}\widetilde{F}^{ab}
\end{equation}
with the conventions
\begin{equation*}
\widetilde{F}^{ab}=\frac{1}{2}\varepsilon ^{abcd}F_{cd}
\end{equation*}
\begin{equation*}
a,b,c....\equiv Tetrad\ \ indexes
\end{equation*}
For the tensor of electromagnetic fields $F$ , we propose the form similar
to the Boyer and Lindquist generalization to the Kerr-Newman problem (
axially symmetric metric ), for example:
\begin{equation*}
F=F_{20}dr\wedge \lbrack dt-a\sin ^{2}\theta d\varphi ]+F_{31}\sin \vartheta
d\theta \wedge \lbrack \left( r^{2}+a^{2}\right) d\varphi -adt]
\end{equation*}
that, in the tetrad system (that does not depend on the explicit form of $%
\omega ^{a}$), we simply have
\begin{equation}
F=F_{20}\omega ^{2}\wedge \omega ^{0}+F_{31}\omega ^{3}\wedge \omega ^{1}
\end{equation}
Let us note that $F_{20}\ $and $F_{31}$ are the only components of the
fields in the tetrad (rotating geometry) and the energy-momentum tensor
takes the diagonal form. Now
\begin{equation*}
\mathcal{L}_{BI}=\frac{b^{2}}{4\pi }\left( 1-\sqrt[2]{1-\frac{2S}{b^{2}}-%
\frac{P^{2}}{b^{4}}}\right) \ \ \ \ \ \ \ \Longrightarrow
\end{equation*}
in the orthonormal frame (tetrad)
\begin{equation*}
S=-\frac{1}{2}\left[ \left( F_{31}\right) ^{2}-\left( F_{02}\right) ^{2}%
\right]
\end{equation*}
\begin{equation*}
P=\frac{1}{2}\left( F_{13}F_{02}\right)
\end{equation*}
\begin{equation*}
\mathcal{L}_{BI}=\frac{b^{2}}{4\pi }\left( 1-\text{\textsc{R}}\right)
\end{equation*}
where we defined
\begin{equation}
\text{\textsc{R}}=\sqrt[2]{1+\frac{\left( F_{13}\right) ^{2}-\left(
F_{02}\right) ^{2}}{b^{2}}-\frac{\left( F_{13}F_{02}\right) }{b^{4}}}=\sqrt[2%
]{\left[ 1+\left( \frac{F_{13}}{b}\right) ^{2}\right] \left[ 1-\left( \frac{%
F_{02}}{b}\right) ^{2}\right] }
\end{equation}
Finally, looking at (45), the components from $T_{b}^{a}$ are
\begin{equation*}
T_{..0}^{0}=\mathcal{L}_{BI}-\frac{\left( F_{20}\right) ^{2}}{4\pi \text{%
\textsc{R}}}-\frac{1}{4\pi \text{\textsc{R}}}\left( \frac{P}{b}\right) ^{2}
\end{equation*}
\begin{equation*}
T_{..1}^{1}=\mathcal{L}_{BI}+\frac{\left( F_{31}\right) ^{2}}{4\pi \text{%
\textsc{R}}}-\frac{1}{4\pi \text{\textsc{R}}}\left( \frac{P}{b}\right) ^{2}
\end{equation*}
\begin{equation*}
T_{..2}^{2}=\mathcal{L}_{BI}-\frac{\left( F_{20}\right) ^{2}}{4\pi \text{%
\textsc{R}}}-\frac{1}{4\pi \text{\textsc{R}}}\left( \frac{P}{b}\right) ^{2}
\end{equation*}
\begin{equation}
T_{..3}^{3}=\mathcal{L}_{BI}+\frac{\left( F_{31}\right) ^{2}}{4\pi \text{%
\textsc{R}}}-\frac{1}{4\pi \text{\textsc{R}}}\left( \frac{P}{b}\right) ^{2}
\end{equation}
From here we can easily see the typical structure of the energy-momentum
tensor of Born Infeld in the tetrad that we used in the analysis of Sect. 4
[8,11,12]
\begin{equation*}
-T_{00}=T_{22}=\frac{b^{2}}{4\pi }\left( 1-\widetilde{u}\right)
\end{equation*}
\begin{equation}
T_{11}=T_{33}=\frac{b^{2}}{4\pi }\left( 1-\widetilde{u}^{-1}\right)
\end{equation}
where we defined the invariant quantity
\begin{equation*}
\widetilde{u}\equiv \sqrt{\frac{(\overline{F}_{31})^{2}+1}{1-\left(
\overline{F}_{02}\right) ^{2}}}
\end{equation*}
with
\begin{equation*}
\overline{F}_{ab}\equiv \frac{F_{ab}}{b}
\end{equation*}

\section{Acknowledgements:}

The author is grateful to Professors Norma S\'{a}nchez, Yu. Stepanovsky,
Mrs. G. Sadukovskaya and the Professor G. Afanasiev. Thanks are also due to
the Directorate of Joint Institute of Nuclear Research (JINR), and all
people of the Bogoliubov Laboratory of Theoretical Physics for their
hospitality and support.

\section{References:}

\bigskip

\ \ \ \ \ [1] R. P. Kerr, Phys. Rev. Letters \textbf{11}, 237 (1963).

[2] E. T. Newman and A. I. Janis, J. Math. Phys. \textbf{6}, 915 (1965).

[3] E. T. Newman, E. Couch, K. Chinnapared, A. Exton, A. Prakash, and R.
Torrence, J. Math. Phys.\textbf{\ 6}, 918 (1965).

[4] A. I. Janis and E. T. Newman, J. Math. Phys. \textbf{6}, 902 (1965).

[5] Hongsu Kim, Phys. Rev. D \textbf{59,} 064002 (1999); M. Ba$\widetilde {%
\text{n}}$ados, C. Teitelboim and J. Zanelli, Phys. Rev. Lett. \textbf{69}%
,1849 (1992).

[6] R. M. Texeira Filho and V. B. Bezerra, Phys. Rev. D \textbf{64, }084009
(2001); M. Barriola and A. Vilenkin, Phys. Rev. Lett. \textbf{63}, 341
(1989).

[7] E. T. Newman, Phys. Rev. D \textbf{65}, 104005 (2002); S. P. Drake and
P. Szekeres, gr-qc/9807001; S. Yazadjiev, Gen. Rel. Grav.\textbf{\ 32}, 2345
(2000).

[8] B. Hoffmann, Phys. Rev. \textbf{47}, 877 (1935).

[9] G. W. Gibbons and D. A. Rasheed, Nucl. Phys. B \textbf{454}, 185 (1995);
G. W. Gibbons and D. A. Rasheed, Nucl. Phys. B \textbf{476}, 515 (1996); D.
Chruscinski, Phys. Rev. D \textbf{62}, 105007 (2000); D. Sorokin,
hep-th/9709190 .

[10] A S. Prudnikov, Yu. Brychov and O. Marichev, \textit{Integrals and
Series} (Gordon and Breach, New York, 1986).

[11] M. Born and L. Infeld, Proc. R. Soc.(London) \textbf{144, }425 (1934).

[12] D. J. Cirilo Lombardo, Problems of Atomic Science and Technology\textbf{%
\ \#6, }pp.71-73, (2001); D.J. Cirilo Lombardo, Master thesis , Universidad
de Buenos Aires, Argentine, 2001.

[13] R. H. Boyer and R. W. Lindquist, J. Math. Phys.\textbf{8}, 265 (1967).

[14] C. Misner, K. Thorne and J. A. Weeler, \textit{Gravitation} ( Freeman
and Company, New York, 1973).

[15] A. Ya. Burinskii, hep-th/0008129; see also in \textit{Noncommutative
Structures in Mathematics and} \textit{Physics}, Edited by S. Duplij and J.
Wess (Kluwer Publishers, The Netherlands, 2001), p.181.

[16] A. Ya. Burinskii, JETP lett. \textbf{39}, 193 (1974).

[17] A. Ya. Burinskii et al., Phys. Rev. D \textbf{65}, 064039 (2002).

[18] B Carter, Phys. Rev. \textbf{174}, 1559 (1968); S. Chandrasekhar,
\textit{The Mathematical Theory of Black Holes} (Oxford University Press,
New York,1992).

[19] H. Salazar, A. Garc\'{i}a and J. Pleb\'{a}nski, J. Math. Phys.\textbf{\
28}, 2171 (1987).

[20] Yu. Stepanovsky, Electro-Magnitie Iavlenia \textbf{3}, Tom 1, 427
(1988).

[21] R. Metsaev and A. Tseytlin, Nucl. Phys. B \textbf{293}, 385 (1987).

\bigskip

\bigskip

\bigskip

\bigskip

\bigskip

\bigskip

\bigskip

\bigskip

\bigskip

\bigskip

\bigskip

\bigskip

\bigskip

\bigskip

\bigskip

\bigskip

\bigskip

\bigskip

\bigskip

\bigskip

\bigskip

\bigskip

\bigskip

\bigskip

\bigskip

\bigskip

\bigskip

\bigskip

\bigskip

\bigskip

\bigskip

\bigskip

\bigskip

\bigskip

\bigskip

\bigskip

\bigskip

\bigskip

\bigskip

\bigskip

\bigskip

\bigskip

\bigskip

\bigskip

\bigskip

\end{document}